\documentstyle[revtex]{aps}

\newcommand{\be}{\begin{equation}}
\newcommand{\ee}{\end{equation}}
\newcommand{\beq}{\begin{eqnarray}}
\newcommand{\eeq}{\end{eqnarray}}

\begin{document}
\draft
%\pagestyle{empty}                                      %%%To be commented
%\centerline{Version:\vday \hfill   NTUTH--93--09}   %%%To be commented
%
\vfill
\begin{title}
Is the Top Quark Really Heavier than the $W$ Boson?
\end{title}
\vfill
\author{Wei-Shu Hou}
\begin{instit}
Department of Physics, National Taiwan University,
Taipei, Taiwan 10764, R.O.C.
\end{instit}
%
%\author{W. K. Sze}
%
%\begin{instit}
%Department of Physics, National Taiwan Normal University,
%Taipei, Taiwan, R.O.C.
%\end{instit}
\receipt{\today}
%
%\vskip -1cm
\vfill
\begin{abstract}
%The $m_t < M_W$ region has not yet been carefully studied.
Scalar induced top decays may drastically suppress $B(t\to \ell\nu + jet)$
and still hide the top below $M_W$.
The $p\bar p$ collider experiments should enlarge the scope
and study the $m_t - B(t\to\ell\nu j)$ plane.
Specific model signatures such as
$t\to ch^0\to cb\bar b$ (multiple high $p_T$ $b$-jets)
and $t\to bH^+\to bc\bar s$, $b\tau^+\nu$ (with $B(t\to b\tau\nu)
\ \raisebox{-.5ex}{\rlap{$\sim$}} \raisebox{.4ex}{$<$}\  1/3$)
should be explored.
Without ruling out these possibilities,
isolated lepton signals in the future
might actually be due to the 4th generation $t^\prime$ or $b^\prime$ quark,
while top quark and toponium physics could still turn up at LEP-II.

\end{abstract}
%%%%%%%%%%%%%%%%%%%%%%%%%% See PRL 69 #24. 1992 for the listings
%\newpage
%
\narrowtext
\pagestyle{plain}

%\section{Introduction}

When the $\tau$-lepton \cite{Perl}
and the $b$-quark \cite{Oops} were discovered in the 1970's,
the top quark was thought to be just around the corner,
with mass of order $10 - 30$ GeV.
Shortly after the ARGUS observation of
large $B^0$--$\bar B^0$ mixing \cite{ARGUS},
however, the prejudice started to shift.
To date, the top remains elusive.
A global fit \cite{PDG} to
LEP and low energy electroweak data
suggests a high value of
 \be
m_t = 150^{\mbox{$+23$}}_{\mbox{$-26$}} \pm 16\ \mbox{GeV}.
\label{eq:tfit}
 \ee
The published direct search limit by the CDF Collaboration is \cite{CDF}
 \be
m_t > 91\ \mbox{GeV}, \label{eq:tCDF}
 \ee
based on %$4$ pb$^{-1}$ of
data collected in
the 1989 Tevatron collider run at Fermilab.
With an order of magnitude more data from the 1993 collider run,
the CDF and D0 collaborations have extended the limit
to about $120$ GeV \cite{TevNew}.
Though not yet conclusive,
a handful of events hint at a top mass
consistent with the global fit result of eq. (\ref{eq:tfit}).
Thus, once again we have the expectation that
the top is ``just around the corner", and
should be discovered in the 1994 Tevatron collider run.

So, the top seems to be very heavy.
But what {\it hard} evidence do we really have?
Eq. (\ref{eq:tfit}) assumes the 3 generation Standard Model (SM).
If the top is in fact relatively light,
it actually suggests the existence of
{\it new} weak doublets (or other multiplets)
with sizable splittings.
%for example, a fourth generation.
Eq. (\ref{eq:tCDF}) assumes the SM branching ratio
 \be
B_{s.l.}\equiv B(t\to \ell^+\nu + X) \simeq \frac{1}{9}. \label{eq:SMBR}
 \ee
%where $\ell = e,\ \mu$.
The limit would weaken if $B_{s.l.}$ is %, for some reason,
far below this value. A weaker limit \cite{UA2} of
% \be
$m_t > 55\ {\rm GeV}$ %, \label{eq:UA2}
% \ee
is obtained from the measured $W$ width.
This limit would soften
if the quark mixing element $\vert V_{tb}\vert < 1$.
The truly model independent limit
comes from $Z\to t\bar t$ search \cite{PDG} at LEP/SLC,
% \be
$m_t \ \raisebox{-.5ex}{\rlap{$\sim$}} \raisebox{.4ex}{$>$}\  M_Z/2$.
%\label{eq:tabs} % \ee
Thus, a relatively light top
should not be a forgone conclusion.

The mass region that needs special scrutiny is
 \be
M_Z/2  \ \raisebox{-.5ex}{\rlap{$\sim$}} \raisebox{.4ex}{$<$}\ m_t
       \ \raisebox{-.5ex}{\rlap{$\sim$}} \raisebox{.4ex}{$<$}\  M_W+m_b.
\label{eq:range}
 \ee
%or, if one takes eq. (\ref{eq:UA2}),
%$55\ \mbox{GeV} \ltap m_t < M_W+m_b$.
The reason is as follows.
For $m_t > M_W + m_b$, $\Gamma(t\to bW)$ is
rather large, and eq. (\ref{eq:SMBR}) should hold.
However, below the $W$ threshold,
$t\to bW^*$ is suppressed both by phase space
(three body) and propagator effects.
If new interactions induce
{\it two body} decays, %of the top quark,
they would be {\it relatively} enhanced, thereby suppressing $B_{s.l.}$.
The new coupling strength should
not be much weaker than $SU_L(2)$ gauge coupling,
and the modes should be relatively obscure such
that they have not yet been studied.
{\it Effectively this can be achieved only by (pseudo)scalar
interactions} (including sfermions).
%the $t\to bW^*$ decay is phase space and propagator suppressed.
%Any potential {\it two-body} final state due to
%new physics would be relatively enhanced,
%resulting in smaller $B_{s.l.}$.
Possible scenarios are:
$t\to cb\bar b$ where $b\bar b$ comes from
the decay of a light neutral scalar boson
%(that mediates FCNC $t\to c$ transitions)
\cite{Hou,HW};
or $t\to bc\bar s$, where $c\bar s$ comes from a
charged Higgs boson \cite{GJ}.
A third possibility of $t\to bH^+\to b\tau^+\nu$ \cite{tbH}
is unlikely \cite{Hewett}
in light of recent limits on $b\to s\gamma$ from CLEO \cite{CLEO}.
The two scenarios lead to $t\to 3$ jets final state,
which is very hard to disentangle in hadronic collisions.
We suggest that the newly accumulated data should
be used to explore the low $m_t$ possibility of eq. (\ref{eq:range}),
by taking $B_{s.l.}$ as a free parameter.
We then comment on %other constraints and
means of detecting the specific, new modes, as well as implications
of having a light top quark.

%\section{Model Independent Search for Light Top}

The semileptonic decays,
except perhaps $t\to b\tau^+\nu$,
are expected to be mediated by the $W$ boson.
%The same cannot be said about $t\to 3$ jets.
Eq. (\ref{eq:SMBR})
assumes %that $t\to bu\bar d$, $bc\bar s$ are
%also mediated purely by $W$ bosons, so
$\Gamma^{SM}(t\to 3\ \mbox{jets}) \simeq
               6\, \Gamma(t\to e^+ \nu + \mbox{jet})$.
However, some new interaction
could enhance the $t\to jets$ mode,
%\begin{equation}
that is $\Gamma_{3j} = \Gamma^{SM}_{3j} + \Gamma^{\prime}_{3j}$,
%\label{eq:t3jNEW}
%\end{equation}
where $\Gamma_{3j}^{\prime}$ is the additional 3-jet width.
Eq. (\ref{eq:SMBR}) gets modified by the factor
 \be
R\equiv B_{s.l.}/B^{SM}_{s.l.}
      = \left(1 - B^{\prime}\right),
%         &\simeq& \frac{1}{9} \left(1 - B^{\prime}\right),
\label{eq:R}
 \ee
where
% \be
$ B^{\prime} \equiv \Gamma^{\prime}/\left(\Gamma^{SM}_{tot}
                              +\Gamma^{\prime}\right)
$ %\label{eq:BRnew}
% \ee
is the new physics branching ratio.
%With eqs. (\ref{eq:R}) and (\ref{eq:BRnew}),
%Written as $\Gamma^{\prime}$,
It could be other new top decay possibilities,
{\it e.g.} %$t\to b\tau\nu$ \cite{tbH} or top decay to
light supersymmetric (SUSY) particles \cite{Baer}.

The  search mode branching ratios become
\begin{equation}
 \frac{4}{81} \left(1 - B^{\prime}\right)^{\mbox{2}},\
 \frac{8}{27} \left(1 - B^{\prime}\right)
              \left(1 + \frac{1}{2} B^{\prime}\right),
\label{eq:12BR}
\end{equation}
for $t\bar t\to \ell_1^+\ell_2^{-} + \nu\nu + jets$ and
$\ell^\pm + \nu + jets$, respectively.
It is clear that if $B^{\prime}\sim 1$ (or $B_{s.l.} \rightarrow 0$),
the dilepton signature and therefore
the limit of eq. (\ref{eq:tCDF}) would rapidly become ineffective.
The weaker limit of $m_t < 77$ GeV \cite{enuj} (assuming SM)
obtained from $\ell^\pm + \nu + jets$ search, is
less sensitive to $B^{\prime}$.

To explore the possibility of $B^{\prime}\sim 1$ ($R\rightarrow 0$),
one should keep $B(t\to \ell^+\nu + X)$
%or $B(t\to jets)$
as a free parameter. %and explore the
%$m_t - B_{s.l.}$ plane.
%To do this,
One has to {\it purposely}
keep the cuts on lepton $p_T$ and missing $E_T$ relatively low,
otherwise the signal events %themselves
might get rejected by stiffer cuts
aimed at searching for a heavier top quark.
Furthermore, as the expected number of signal events dwindle,
a more careful study is needed to suppress background
to a level below what has been achieved in the analysis of 1989 data.
To the best of our knowledge,
this has not yet been done by the experimental collaborations \cite{UA1}.

Let us take theoretical ${t\bar t}$ cross sections and use eq. (\ref{eq:12BR})
to scale the $m_t$ limits from 1989 CDF data as function of $R$.
The results are given in Fig. 1, where we have assumed
constant efficiencies and acceptance.
Clearly, for smaller $B_{s.l.}$,
only the experimental groups can give definite curves,
but they should in general fall below those shown in Fig. 1.
Newer data from 1993 and later runs can be used to extend
the excluded domain in the $m_t - B_{s.l.}$ plane.
Note that for smaller $R$, the single lepton signal is more effective.
However, {\it both methods fail in the vicinity of $R\cong 0$},
since the signal would vanish against background.

%\section{Specific Models and Direct Search}

We turn to specific models
that may allow $m_t$ to fall in the range of eq. (\ref{eq:range}).
The first mode is $t\to ch^0$ followed by
$h^0\to b\bar b$ \cite{Hou,HW}, within the context of
two Higgs doublet models (2HDM).
At first sight this seems absurd,
since in standard types of 2HDM, just like in SM,
tree level FCNC Higgs couplings are absent by construction \cite{GW}.
However, this turned out \cite{CS} to be an overkill.
%and basically reflects our lack of knowledge of
%quark mixing in the 1970's.
{\it Neutral Higgs bosons can have
flavor changing neutral couplings (FCNC)  %(extended definition of FCNC)
$\lambda_{ij}$      of order  $\sqrt{2m_i m_j}/v$, and
with normal Higgs boson masses}
of order the vacuum expectation value $v$.
%Low energy FCNC mediated by neutral Higgs bosons
%are naturally suppressed by light external fermion masses
%in these types of 2HDMs \cite{CS},
%but
The $t$-$c$-$h^0$ coupling $\lambda_{tc}$
is precisely the largest \cite{Hou}.
%It was recently pointed out \cite{CHK}
%that a two loop mechanism implies the limit $m_t \gtap 200$ GeV
%from $\mu\to e\gamma$, making $t\to ch^0$ implausible.
%However,
It is furthermore possible that
only $u$-type quarks have FCNC couplings,
while neutral Higgs couplings to $d$-type quarks and
charged leptons are diagonal as in SM and standard 2HDM.
In this variant \cite{Hou}, stringent limits from
$\mu\to e\gamma$ \cite{CHK}
and $K^0$--$\bar K^0$ mixing, {\it etc.}, are evaded,
while limits from $D^0$--$\bar D^0$ mixing are rather forgiving,
so $m_{h^0}$ and/or $\lambda_{tc}$
are practically unconstrained.
We remark that, within the
context of {\it general} 2HDM,
%whether there is tree level
%FCNC Higgs couplings or not,
neutral Higgs boson mass limits from LEP \cite{PDG} are weakened,
and $m_{h^0} < M_Z/2$
is still allowed.

We assume $h^0\to b\bar b$ with approximately SM width and explore
 \be
m_{h^0} < m_t - 10\ \mbox{GeV},\
\lambda_{ct} \geq \sqrt{2m_c m_t}/v, \label{eq:tchpar}
 \ee
The $t\to cb\bar b$ \cite{Hou,HW} final state is purely hadronic,
and is very suppressed in SM.
For both scalar and pseudoscalar $h^0$,
for the range of eq. (\ref{eq:tchpar}),
$\Gamma^{\prime} = \Gamma(t\to ch^0)$
% \be
%  \frac{\lambda_{ct}^2}{32\pi}\, m_t
%                            \left(1-\frac{m_{h^0}^2}{m_t^2}\right)^2
%                           \geq \frac{0.049\, m_c\, m_t^2}{16\pi v^2},
%\label{eq:tch}
% \ee
$\cong (\lambda_{ct}^2/32\pi)\, m_t\,
                            (1-{m_{h^0}^2}/{m_t^2})^2$
                           $\geq 0.049\, m_c\, m_t^2/16\pi v^2$.
%couplings since $m_c/m_t \ll 1$,
%Taking into account propagator effects \cite{BDKKZ}
Together with $\Gamma_{tot}^{SM} = \Gamma(t\to bW^*)$,
$B^{\prime}$ and $B_{s.l.}$ can be readily estimated from
eq. (\ref{eq:R}). % and (\ref{eq:BRnew}).
Note that $B^\prime = 1-R$ is the $t\to ch^0$ branching ratio.
We plot $m_t$ {\it vs.} $R$ in Fig. 2 for various
parameters satisfying eq. (\ref{eq:tchpar}).
Compared with Fig. 1, it is clear that
a large parameter range is allowed,
especially for light $m_{h^0}$ and large $\lambda_{ct}$.

The second mode is $t\to bH^+$ followed by $H^+\to c\bar s$
\cite{GJ}.
%It should be remarked that, in SUSY type of 2HDM's, the
%mode $t\to b\tau^+\nu$, mediated by charged Higgs bosons,
%can get quite enhanced.
We remark that recent data \cite{CLEO}
on $b\to s\gamma$ and $B\to K^*\gamma$ are in good agreement
with SM expectations, which implies \cite{Hewett} that
$m_{H^+} > m_t$ in SUSY type of 2HDM.
This rules out the possibility of
$B(t\to b\tau^+\nu) \to 1$.
However, in {\it non-SUSY type of 2HDM},
where $t$--$b$--$H^+$ coupling is of the form
% \be
$\frac{\sqrt{2}}{v}\, V_{tb}\, \cot\beta\,
     \bar t \left(m_t L - m_b R\right) b + h.c.$,
% \ee
one obtains the rough limit $\tan\beta
\ \raisebox{-.5ex}{\rlap{$\sim$}} \raisebox{.4ex}{$<$}\ 0.5$
for light $m_{H^+}$ \cite{Hewett}.
(Note that one could also have charged Higgs bosons from ``nonstandard"
2HDM's that possess FCNC Higgs couplings.)
We plot $m_t$ {\it vs.} $R$ in Fig. 3
for various parameters satisfying
% \beq
$41\ \mbox{GeV} < m_{H^+} < m_t - 10\ \mbox{GeV}$ and
$\tan\beta > 0.5.$ %\label{eq:tbHpar}
% \eeq
%It is clear that the suppression factor $R$ is strongly dependent
%on $\tan\beta$.
Note that since $H^+$ couplings
share a common $\cot\beta$,
the relative rate for $H^+\to c\bar s$ {\it vs.} $\tau^+\nu$
is roughly $3\, m_c^2\ :\ m_\tau^2$.

As noted earlier, $1-R$ is the yet unobserved $t\to ch^0$ or
$t\to bH^+$ branching ratio.
As $R\rightarrow 0$, so $B^{\prime} \to 1$,
one should look for {\it direct} observables from these decay modes.
For $t\to cb\bar b$, both $b$ jets should be harder than the single
$b$-jet from $t\to bW^*$ decay.
This is because the virtual $W$ tends
to be as close to mass shell as possible.
Hence, one possible way to identify $t\to ch^0 \to cb\bar b$
in case it predominates is to tag for (relatively) high
$p_T$ multiple $b$-jets \cite{Mueller}.
%Assuming tagging efficiency of order 0.2 and
%with integrated luminosity of 10 pb$^{-1}$,
%demanding 3 or more $b$-jets with $p_T$ greater than,
%say 15 GeV, should lead to some 50 events or more.
%Background calculations for multiple high $p_T$ $b$-jets,
%however, are lacking.
%
%
For the $t\to bH^+$ mode, clearly $t\to bc\bar s$ would be
difficult to disentangle from multijet background,
unless
$b$ tagging {\it plus} charm tagging can work together rather well.
The better hope is to utilize the $t\to b\tau^+\nu$ mode, which
accounts for $1/3 - 1/4$ of $t\to bH^+$ transitions in non-SUSY type
of 2HDM's. Thus, CDF and D0 should continue the
$t\to bH^+\to b\tau^+\nu$ search of UA1/UA2 \cite{tbH}
for the mass region of eq. (\ref{eq:range}),
but allowing $B(t\to b\tau\nu)
\ \raisebox{-.5ex}{\rlap{$\sim$}} \raisebox{.4ex}{$<$}\ 1/3$.

%\section{Discussion and Summary}

Some discussion is in order.
First, $m_t$ in the range of eq. (\ref{eq:range})
is more ``normal" since $m_t/m_b \sim m_c/m_s$.
However, the global fit of eq. (1) now implies \cite{PDG}
% \be
$m_t^2 + \sum_i (c_i/3) \Delta m_i^2 < \left(194\ \mbox{GeV}\right)^2$,
% \ee
where $c_i = 1(3)$ for color singlets(triplets),
and $\Delta m_i^2 \geq (m_1-m_2)^2$ is the splitting
in {\it new} weak doublets.
If $\Delta m_i^2$ comes solely from the extra Higgs doublet,
$\vert m_{H^+} - m_{h^0}\vert$ should be of order 300 GeV.
It may be more plausible to have
a fourth generation with $\vert m_{t^\prime} - m_{b^\prime}\vert
\ \raisebox{-.5ex}{\rlap{$\sim$}} \raisebox{.4ex}{$<$}\  150$ GeV.
A heavy, fourth generation is
favored from the point of view of
dynamical symmetry breaking \cite{dsb}.
The new neutral heavy lepton, however, has to be heavier than
$M_Z/2$ to satisfy neutrino counting in $Z$ decay \cite{PDG},
which is itself an interesting situation.
Although the $b^\prime$ quark may also be
obscured by scalar induced decay
(or loop-induced FCNC decay \cite{bp}),
$t^\prime$ decay should
be dominated by $t^\prime \to (b,\ b^\prime) + W$,
where $W$ is on-shell ($B(t^\prime\to th^0)$ should be less than $1/2$).
Thus, {\it even if a ``top"-like signal
(isolated $\ell^\pm +$ missing $E_T$)
is discovered at the Tevatron,
it may well be due to $t^\prime$ (or $b^\prime$) rather than $t$},
and much work would be needed
to clarify the actual flavor involved!

Second, working along the lines sketched in Fig. 1, with diligence
and luck a ``light top" may surface at the Tevatron.
If not, it is reassuring that the mass range of eq. (\ref{eq:range})
can be fully covered by LEP-II as it turns on in 1997.
It would be amusing that not only we would see
the crossing of $e^+e^- \to t\bar t$
(and perhaps $b^\prime\bar b^\prime$) new flavor threshold,
we would actually be able to study {\it toponium} physics
at LEP-II {\it afterall}. The toponium width would
be dominated by single top (scalar induced) decay,
but the spectrum would be retained, with rich phenomenology \cite{KZ}.
Third,
with extra Higgs doublets and (most likely) new fermion generations,
$B^0$-$\bar B^0$ mixing can easily be accommodated.
Furthermore, $B_s$-$\bar B_s$ mixing is no longer
necessarily close to maximal. All considerations
regarding $CP$ violation in $B$ sector are enriched,
{\it e.g.} the unitarity triangle no longer closes.

In summary,
new physics due to light scalar particles
that preferentially couple to the top, but decay
into rather elusive final states,
could suppress $B(t\to \ell\nu + j)$
and thus allow for $m_t < M_W + m_b$,
evading the Tevatron bound.
 %since hard, isolated leptons from top decay may be
 %very suppressed or absent.
The experiments should therefore %loosen $B(t\to e\nu j) \simeq 1/9$ and
explore the $m_t - B_{s.l.}$ plane.
In addition, the process $t\to ch^0\to cb\bar b$ may
be searched for by tagging multiple high $p_T$ $b$-jets,
while $t\to bH^+\to bc\bar s,\ \tau^+\nu$
can be studied by searching for $b\tau\nu$,
but assuming $B(t\to b\tau\nu)
\ \raisebox{-.5ex}{\rlap{$\sim$}} \raisebox{.4ex}{$<$}\ 1/3$.
The standard electroweak fit result of $m_t \sim 150$ GeV would imply
large splittings in other weak doublets, such as a 4th generation.
Hence, it may be the $t^\prime$ quark that
gets discovered at Tevatron in 1994.
If $m_t > M_W$ cannot be demonstrated at the Tevatron beyond doubt,
the top may show up at LEP-II.
If the scenario is realized,
there would be {\it very rich} phenomena unfolding in the near future.
After being elusive for 15 years,
the top quark may surprise us once again.

\acknowledgments
We thank W. K. Sze for participation at an earlier
stage of this work,
and J. Mueller and G. P. Yeh for discussions
on experimental questions.
This work is supported in part by grant NSC 82-0208-M-002-151
of the Republic of China.

\vskip -1cm
\figure{Schematic limit of $m_t$ {\it vs.}
$R\equiv B_{s.l.}/B_{s.l.}^{SM}$ for
dilepton ($+$) and single lepton ($*$) signals,
scaled from 1989 CDF data. %, assuming constant efficiencies and acceptance.
The region below the curve is ruled out.}

\vskip -1cm
\figure{Effect of $t\to ch^0$ mode.
The dotted, dashed and solid curves are for
$m_{h^0} = m_t - 10$, $20$, $30$ GeV, respectively,
while each set (from below) corresponds to
$\lambda_{ct}v/\sqrt{2m_cm_t} = 1$, $2$, $4$.}%

\vskip -1cm
\figure{Effect of $t\to bH^+$ mode.
The dotted, dashed and solid curves are for
$41$ GeV $< m_{H^+} = m_t - 10$, $20$, $30$ GeV, respectively,
while each set (from below) corresponds to
$\cot\beta = 0.5$, $1$, $2$.}
\end{document}